\documentstyle[preprint,eqsecnum,aps]{revtex}
\begin{document}
\def\appls{\hbox{$<$\kern-.75em\lower 1.00ex\hbox{$\sim$}}}
\draft
\title{UNITARITY AND INTERFERING RESONANCES IN\\
$\pi\pi$ SCATTERING AND IN PION PRODUCTION $\pi N \to \pi\pi N$.}
\author{M. Svec\footnote{electronic address:  
milo@smetana.physics.mcgill.ca}}
\address{Physics Department, Dawson College, Montreal, Quebec,  
Canada H3Z 1A4\\
and\\
Physics Department, McGill University, Montreal, Quebec, Canada H3Z 1A4}
\maketitle
\begin{abstract}
Additivity of Breit-Wigner phases has been proposed to describe  
interfering resonances in partial waves in $\pi\pi$ scattering. This  
assumption leads to an expression for partial wave amplitudes that  
involves products of Breit-Wigner amplitudes. We show that this  
expression is equivalent to a coherent sum of Breit-Wigner  
amplitudes with specific complex coefficients which depend on the  
resonance parameters of all contributing resonances. We use  
analyticity of $\pi\pi$ partial wave amplitudes to show that they  
must have the form of a coherent sum of Breit-Wigner amplitudes with  
complex coefficients and a complex coherent background. The  
assumption of additivity of Breit-Wigner phases is a new constraint  
on partial wave amplitudes independent of partial wave unitarity. It  
restricts the partial waves to analytical functions with very  
specific form of residues of Breit-Wigner poles. Since there is no  
physical reason for such a restriction, we argue that the general  
form provided by the analyticity is more appropriate in fits to data  
to determine resonance parameters. The partial wave unitarity can  
be imposed using the modern methods of constrained optimization. We  
discuss the production amplitudes in $\pi N\to\pi\pi N$ reactions  
and use analyticity in the dipion mass variable to justify the  
common practice of writing the production amplitudes in production  
processes as a coherent sum of Breit-Wigner amplitudes with free  
complex coefficients and a complex coherent background in fits to  
mass spectra with interfering resonances. The unitarity constraints  
on $\pi\pi$ partial wave amplitudes with resonances determined from  
fits to mass spectra of production amplitudes measured in $\pi N \to  
\pi\pi N$ reactions can be satisfied with an appropriate choice of  
complex residues of contributing Breit-Wigner poles.
\end{abstract}

\section{Introduction}

In 1930's, Breit and Wigner introduced\cite{wigner33,wigner36} a  
parametrization of resonances observed in the energy dependence of  
integrated and differential cross-sections of nuclear reactions. The  
original Breit-Wigner formula was only a one-resonance  
approximation and its justification was initially only  
phenomenological. A theoretical justification for Breit-Wigner  
formula later emerged from quantum collision theory\cite{blatt52}.  
The evident existence of multiple and overlapping resonances in  
nuclear reactions led to two distinct generalizations of the  
Breit-Wigner formula for an isolated resonance to multiresonance  
description of the scattering process.

One generalization was undertaken by  
Feshbach\cite{feshbach58,feshbach67}, Humblet\cite{humblet67} and  
McVoy\cite{mcvoy67} who used the analyticity properties of the  
$S$-matrix to show that the transition matrix can be written as a  
coherent sum of Breit-Wigner terms with complex coefficients and a  
coherent background. Since the transition matrix must satisfy  
unitarity, the parameters and coefficients of this multiresonance  
parametrization are not independent\cite{feshbach67,rosenfeld70}. In  
principle it is possible to use the methods of nonlinear  
programming\cite{luenberger73,gill81} and constrained optimization  
with computer programs such as MINOS developed at Stanford  
University\cite{murtagh83} to impose the conditions of unitarity in  
fitting the experimental data.

Another approach to multiresonance description of scattering  
process was proposed by Hu in 1948\cite{hu48}. He observed that the  
Breit-Wigner contribution of an isolated resonance to the $S$-matrix  
is unitary and proposed to describe the multiresonance  
contributions in the $S$-matrix by the product of isolated  
Breit-Wigner contributions for each resonance. Since each term is  
unitary, the product also satisfies unitarity. The partial wave  
phase shift is then a sum of Breit-Wigner phases of contributing  
resonances and a background phase. As a result, the expressions for  
partial wave amplitudes involve products of Breit-Wigner amplitudes.  
This method has been recently used by Bugg et al\cite{bugg96} and  
by Ishida et al\cite{ishida96} in their analyses of $\pi\pi$ phase  
shift data.

Up to now the connection between these two descriptions of  
multiresonance contributions (interfering resonances) has not been  
clarified. In this work we show that the Hu description is a special  
case of a more general description based on analyticity. We show  
that the Hu method also leads to a coherent sum of Breit-Wigner  
amplitudes with complex coefficients and a complex coherent  
background for any partial wave as expected from the analyticity of  
the $S$-matrix. However, the complex coefficients have a very  
specific form in terms of resonance parameters of all contributing  
resonances. The assumption of additivity of Breit-Wigner phases is a  
new constraint that restricts the partial waves to analytical  
functions with these specific residues of Breit-Wigner poles.  
Furthermore we show that the additivity of Breit-Wigner phases is an  
assumption entirely independent of the unitarity property of  
partial wave amplitudes which is a condition imposed on their  
inelasticity.

Since there is no physical reason why the physical partial waves  
must have the form of a coherent sum of Breit-Wigner amplitudes with  
the specific complex coefficients required by the additivity of  
Breit-Wigner phases, we conclude that the general form imposed by  
the analyticity is more appropriate for fits to data to determine  
resonance parameters. This conclusion is particularly relevent for  
analysis of interfering resonances in the mass spectra in production  
processes such as $\pi N \to \pi\pi N$ or $pp \to \pi\pi pp$. Using  
analyticity in the invariant mass variables we justify the common  
practice of parametrizing the production amplitudes in terms of a  
coherent sum Breit-Wigner amplitudes with free complex coefficients  
and a complex coherent  
background\cite{lassila67,barnham70,svec97,svec98,alde97,barberis99,bella99,alde98,aitala01}.

The paper is organized as follows. In Section II we briefly review  
the unitarity and the problem of interfering resonances in potential  
scattering since it motivates the analysis in hadronic reactions.  
In Section III we review the two-body partial wave unitarity in  
$\pi\pi$ scattering and its relation to the general form of isospin  
partial waves. In Section IV, we introduce the assumption of  
additivity of Breit-Wigner phases in the $\pi\pi$ scattering and  
show that it leads to partial waves in a form of a coherent sum of  
Breit-Wigner amplitudes with specific complex coefficients and a  
coherent background. In Section V we generalize dispersion relations  
for partial wave amplitudes in $\pi\pi$ scattering to Breit-Wigner  
poles and show that the form obtained from the additivity of  
Breit-Wigner phases is a special case. In Section VI we focus the  
discussion of the two methods to a finite energy interval and argue  
that the addition of Breit-Wigner phases imposes an unjustified  
constraint on fits to data. In Section VII we formulate unitarity  
for production amplitudes in $\pi^-p \to \pi^- \pi^+ n$ reaction and  
contrast it with partial wave unitarity in $\pi\pi$ scattering. In  
Section VIII we show that the method of addition of Breit-Wigner  
phases can be generalized to production amplitudes. We also use  
analyticity in the invariant mass to obtain a more general form for  
production amplitudes in terms of a coherent sum of Breit-Wigner  
amplitudes with free complex coefficients (pole residues) and a  
complex coherent background. We argue that this general form is more  
appropriate in fits to measured mass spectra. Although the  
discussion is confined to pion production amplitudes in $\pi N \to  
\pi\pi N$, the conclusions have general validity. We also comment on  
determination of $\pi\pi$ partial wave amplitudes from resonance  
parameters determined in measurements of production amplitudes in  
$\pi N \to \pi\pi N$ reactions. The paper closes with a summary in  
Section IX.

\section{Unitarity and interfering resonances in potential scattering.}

\subsection{Unitarity}

We will consider the scattering of a spinless particle of mass $m$  
by a real, central potential $V(r)$\cite{joach87}. In the asymptotic  
form of the stationary scattering wave function, the outgoing wave  
is characterized by the scattering amplitude $f(k, \theta)$ where  
$k$ is the wave number of the particle related to its energy by

\begin{equation}
E = {{p^2}\over{2m}} = {{\hbar^2 k^2}\over{2m}}
\end{equation}

\noindent
and $\theta$ is the scattering angle. In the units $\hbar = 1$ the  
wave number $k$ has the meaning of momentum $p$. The scattering  
amplitude can be written in the form
\begin{equation}
f(k,\theta) = \sum\limits^\infty_{\ell = 0} (2\ell + 1) T_\ell (k)  
P_\ell (\cos\theta)
\end{equation}
\noindent
The partial wave amplitudes $T_\ell$ are given by
\begin{equation}
T_\ell = {1\over{2ik}} [S_\ell (k) - 1]
\end{equation}
\noindent
where $S_\ell(k)$ is called $S$-matrix. For elastic scattering
\begin{equation}
S_\ell = e^{2i\delta_\ell (k)}
\end{equation}
\noindent
where the phase-shifts $\delta_\ell$ describe the interaction and  
are related to the potential $V(r)$. For elastic scattering  
$|S_\ell| = 1$ which is the condition of elastic unitarity.

When a particle collides with a target, non-elastic processes are  
possible and particles are removed from the incident (elastic)  
channel. Since the interaction can alter only the outgoing part of  
the wave function, we require that the amplitude of the outgoing  
wave be reduced if non-elastic processes occur. The reduction of  
scattering amplitudes leads to conditions of inelastic unitarity
\begin{equation}
|S_\ell| \le 1
\end{equation}
\noindent
This suggests that we write
\begin{equation}
S_\ell = \eta_\ell (k) e^{2i\delta_\ell (k)}
\end{equation}
\noindent
where $\eta_\ell$ is called inelasticity and has values
\begin{equation}
0 < \eta_\ell \le 1
\end{equation}
\noindent
The partial wave then has a general form
\begin{equation}
T_\ell = {1\over{2ik}} [\eta_\ell e^{2i\delta_\ell} - 1]
\end{equation}
\noindent
>From (2.8) it follows that
\begin{equation}
Im T_\ell = k |T_\ell|^2 + {1\over{4k}} (1-\eta^2_\ell)
\end{equation}
\noindent
This equation expresses the unitarity condition on the partial  
waves $T_\ell$.

\subsection{Interfering Resonances.}

In the following we will work with partial wave amplitudes
\begin{equation}
t_\ell = kT_\ell = {1\over{2i}} [\eta_\ell e^{2i\delta_\ell} - 1]
\end{equation}
\noindent
and the energy $E$ instead of $k$. A detailed study of the  
potential scattering\cite{joach87} shows that the phase shift may be  
decomposed as $\delta_\ell = \xi_\ell + \rho_\ell$ where $\xi_\ell$  
is the background phase which does not depend on the shape and  
depth of the interaction potential $V(r)$ while the part $\rho_\ell$  
does depend on the details of the potential. Near resonant energy  
$E_r$
\begin{equation}
\tan \rho_\ell \approx {{\Gamma (E)}\over{2(E_r - E)}}
\end{equation}
\noindent
where $\Gamma (E)$ is the width of the resonance. We introduce a  
Breit-Wigner resonance phase $\delta^r_\ell$
\begin{equation}
\delta^r_\ell = \tan^{-1} \{ {{\Gamma (E)}\over{2(E_r - E)}} \} =  
\arg [E_r - E + i{1\over 2} \Gamma (E)]
\end{equation}
\noindent
such that in the energy interval $\Delta E$ centered about $E_r$ we  
have $\rho_\ell \approx \delta^r_\ell$ and
\begin{equation}
\delta_\ell \approx \xi_\ell + \delta^r_\ell
\end{equation}
\noindent
From (2.12) it follows that
\begin{equation}
e^{2i\delta^r_\ell} = {{E - E_r - {1\over 2} i\Gamma}\over{E - E_r  
+ {1\over 2} i\Gamma}} = 1 + 2i a_\ell
\end{equation}
\noindent
where
\begin{equation}
a_\ell = {{-{1\over 2} \Gamma}\over{E-E_r + {1\over 2} i\Gamma}}
\end{equation}
\noindent
is the Breit-Wigner amplitude of the resonance $r$. For an isolated  
resonance we then obtain
\begin{equation}
t_\ell = {1\over{2i}} ( \eta_\ell e^{2i\xi_\ell} - 1 ) + \eta_\ell  
e^{2i\xi_\ell} ({{-{1\over 2} \Gamma}\over{E - E_r + {1\over 2}  
i\Gamma}})
\end{equation}
\noindent
If $N$ resonances contribute over an interval $\Delta E$ then,  
following Hu\cite{hu48} and references\cite{bugg96,ishida96}, we can  
write
\begin{equation}
e^{2i\delta^r_\ell} = \prod\limits^N_{n=1} {{E - E^{(n)}_r -  
{1\over 2} i \Gamma^{(n)}}\over{E - E^{(n)}_r + {1\over 2} i  
\Gamma^{(n)}}} = \prod\limits^N_{n=1} (1 + 2i a^{(n)}_\ell)
\end{equation}
\noindent
The prescription (2.17) clearly satisfies unitarity but seems to  
lead to a complicated expression for partial waves $t_\ell$ in terms  
of Breit-Wigner amplitudes $a^{(n)}_\ell$.

On the other hand, analyticity of $S$-matrix was used by  
Feshbach\cite{feshbach58,feshbach67}, Humblet\cite{humblet67} and  
McVoy\cite{mcvoy67} to derive a general form for  
$t_\ell$\cite{feshbach67}
\begin{equation}
t_\ell = B_\ell (E) + \sum\limits^N_{n=1} {{A^{(n)}_\ell  
(E)}\over{E - E^{(n)}_r + {1\over 2} i \Gamma^{(n)} (E)}}
\end{equation}
\noindent
where $B_\ell$ is a background term and $A^{(n)}_\ell (E)$ are  
complex coefficients. The sum in (2.18) can be written as a coherent  
sum of Breit-Wigner amplitudes
\begin{equation}
t_\ell (E) = B_\ell (E) + \sum\limits^N_{n=1} R^{(n)}_\ell (E)  
{{-{1\over 2} \Gamma^{(n)} (E)}\over{E - E^{(n)}_r + {1\over 2} i  
\Gamma^{(n)} (E)}}
\end{equation}
\noindent
In Section IV we show that the prescription (2.17) leads to the  
analytical form (2.19) with specific expressions for the  
coefficients $R^{(n)}_\ell$ and the background $B_\ell$.

\section{Isospin amplitudes and unitarity in $\pi\pi \to \pi\pi$  
scattering.}

Hadron resonances have definite values of spin and isospin. It is  
therefore necessary to express the amplitudes for charged pion  
processes $\pi\pi \to \pi\pi$ in terms of isospin amplitudes $T^I  
(E,\theta)$ with definite isospin $I=0,1,2$ and work with partial  
wave amplitudes $T^I_\ell (E)$\cite{martin76}. At first we will work  
with the center-of-mass energy $E=\sqrt s$ to pursue the analogy  
with the potential scattering.

The partial wave amplitudes $T^I_\ell$ satisfy partial wave  
unitarity equations\cite{martin76,pilk67}
\begin{equation}
Im T^I_\ell = q |T^I_\ell|^2 + \Delta^I_\ell
\end{equation}
\noindent
where $\Delta^I_\ell$ are the contributions from inelastic  
channels, such as $\pi\pi\to\pi\pi\pi\pi$, $K\overline K$,  
$N\overline N$, and $q = \sqrt{{1\over 4} (s-4\mu^2)}$ is c.m.  
momentum where $\mu$ is the pion mass.. Let us write $\Delta^I_\ell$  
in the form
\begin{equation}
\Delta^I_\ell = {1\over{4q}} (1- (\eta^I_\ell)^2)
\end{equation}
\noindent
Then the unitarity equation (3.1) has the same form as (2.9) and  
the partial waves $T^I_\ell$ can be written as
\begin{equation}
T^I_\ell = {1\over{2iq}} \{ \eta^I_\ell e^{2i\delta^I_\ell} -1\}
\end{equation}
\noindent
where the $\delta^I_\ell (E)$ are phase shifts and the inelasticity
\begin{equation}
\eta^I_\ell (E) = \sqrt{1-4q \Delta^I_\ell (E)}
\end{equation}
\noindent
is given by the inelastic unitarity contributions $\Delta^I_\ell$.  
In analogy with potential scattering we expect that $0 < \eta^I_\ell  
\le 1$. As we shall see later, the descriptions of interfering  
resonances in $\pi\pi$ scattering do not depend on the condition  
that $\eta^I_\ell \le 1$.

The positivity of inelasticity $\eta^I_\ell$ in (3.4) imposes a  
constraint
\begin{equation}
\Delta^I_\ell \le {1\over{4q}}
\end{equation}
\noindent
We can now show that the unitarity equation (3.1) admits no  
solution for $\Delta^I_\ell > 1/4q$. If the inelastic unitarity  
contributions satisfy this condition we can write
\begin{equation}
\Delta^I_\ell = {1\over{4q}} (1 + (\eta^I_\ell)^2)
\end{equation}
\noindent
where $(\eta^I_\ell)^2 > 0$. Setting $T^I_\ell = V^I_\ell/2q$ we  
get from the unitarity equation (3.1)
\begin{equation}
(Re V^I_\ell)^2 + (Im V^I_\ell - 1)^2 = - (\eta^I_\ell)^2
\end{equation}
\noindent
which is not possible. Thus the conditions (3.5) represent genuine  
constraints on the inelastic unitarity contributions $\Delta^I_\ell$  
and the parameterization (3.3) of partial wave amplitudes with  
(3.4) is the most general solution of the unitarity equation (3.1)  
for all $E$.

Since the values of inelastic terms $\Delta^I_\ell$ in the partial  
wave unitarity equations (3.1) are not known, we constrain the  
partial waves $T^I_\ell$ by inequalities imposed by the unitarity.  
From the positivity of $\eta^I_\ell$ and the condition (3.5) we  
obtain
\begin{equation}
Im T^I_\ell \le q |T^I_\ell|^2 + {1\over{4q}}
\end{equation}
\noindent
If we add the requirement that $\eta^I_\ell \le 1$, then  
$\Delta^I_\ell \ge 0$ from (3.2), and we also have the usual  
unitarity constraint
\begin{equation}
Im T^I_\ell \ge q |T^I_\ell|^2
\end{equation}
\noindent
The inequality (3.9) implies positivity
\begin{equation}
Im T^I_\ell \ge 0
\end{equation}
\noindent
at all energies.

Finally we note the following observation. Let $f(z)$ be any  
complex function. Then $1+2if(z)$ is a complex function that can be  
written as
\begin{equation}
1 + 2i f(z) = \eta (z) e^{2i\delta(z)}
\end{equation}
\noindent
where $\eta > 0$ and $\delta$ is real. Thus any complex function  
can be written in the form
\begin{equation}
f(z) = {1\over{2i}} (\eta e^{2i\delta} - 1)
\end{equation}
\noindent
and satisfies the equation
\begin{equation}
Im f = |f|^2 + {1\over 4} (1-\eta^2)
\end{equation}
\noindent
We see that the unitarity equations (3.1) are a special case of  
(3.13) with $\eta$ given by (3.4).

\section{Interfering resonances in $\pi\pi$ scattering using the  
addition of Breit-Wigner phases.}

The general form of phase shift parametrization of partial wave  
amplitudes $T^I_\ell$ is
\begin{equation}
T^I_\ell = {1\over{2iq}} [ \eta^I_\ell e^{2i\delta^I_\ell} - 1]
\end{equation}
\noindent
with inelasticity $\eta^I_\ell$ determined by unitarity via (3.4).  
In the following we will omit the indices $\ell$ and $I$ for  
simplicity. In analogy with potential scattering, we decompose the  
phase shifts $\delta$ into two parts
\begin{equation}
\delta = \xi + \delta^r
\end{equation}
\noindent
where $\xi$ is the nonresonant background phase and $\delta^r$ is  
the phase due to physical particle resonances occurring in the  
partial wave $T^I_\ell$. The phase of a single isolated resonance is  
given by the Breit-Wigner formula
\begin{equation}
e^{2i\delta^r} = {{E- E_r - {1\over 2} i \Gamma(E)}\over{E - E_r +  
{1\over 2} i \Gamma (E)}}
\end{equation}
\noindent
Let us consider that $N$ resonances contribute to the partial wave  
amplitude $T^I_\ell$. Following ref.~[12--14] we now assume, that  
the resonant phase shift $\delta^r$ is given by the sum of the  
Breit-Wigner phases of the contributing resonances
\begin{equation}
\delta^r = \sum\limits^N_{n=1} \delta^r_n
\end{equation}
\noindent
We assume that $N$ is finite. Then
\begin{equation}
e^{2i\delta^r} = \prod\limits^N_{n=1} e^{2i\delta^r_n} =  
\prod\limits^N_{n=1} {{E - E_n - {1\over 2} i\Gamma_n}\over{E - E_n  
+ {1\over 2} i \Gamma_n}}
\end{equation}
\noindent
We can write for each Breit-Wigner phase
\begin{equation}
e^{2i\delta^r_n} = 1 + 2i a_n
\end{equation}
\noindent
where $a_n$ is the Breit-Wigner amplitude
\begin{equation}
a_n = {{-{1\over 2} \Gamma_n}\over{E - E_n + {1\over 2} i \Gamma_n}}
\end{equation}
\noindent
Then we can write
\begin{equation}
e^{2i\delta^r} = 1 + 2i T_{res}
\end{equation}
\noindent
where $T_{res}$ is given in terms of products of Breit-Wigner  
amplitudes $a_n$ The partial wave amplitude $T^I_\ell$ then has a  
general form
\begin{equation}
T = {1\over{2iq}} (\eta e^{2i\xi} - 1) + {1\over q} e^{2i\xi} \eta  
T_{res}
\end{equation}

Let us consider the case $N=2$. Then the resonant part of the  
amplitude $T^I_\ell$ is
\begin{equation}
T_{res} = a_1 + a_2 + 2i a_1 a_2
\end{equation}
\noindent
where the interference term
\begin{equation}
a_1 a_2 = {{(-{1\over 2} \Gamma_1)(-{1\over 2} \Gamma_2)}\over{(E -  
E_1 + {1\over 2} i \Gamma_1)(E-E_2 + {1\over 2} i\Gamma_2)}}
\end{equation}
\noindent
With a notation
\begin{equation}
z_k = E_k - {1\over 2} i \Gamma_k
\end{equation}
\noindent
we write
\begin{equation}
{1\over{(E-z_1)(E-z_2)}} = [ {A\over{E-z_1}} + {B\over{E-z_2}} ]  
{1\over C}
\end{equation}
\noindent
The requirement that this equality holds leads to relation
\begin{equation}
(A+B) E - (A z_2 + B z_1) = C
\end{equation}
\noindent
Next we require that $A = - B$ to eliminate the $E$ dependent term,  
and get $C = A (z_1 - z_2)$.
\noindent
Then (4.13) has the form of a sum
\begin{equation}
{1\over{(E - z_1)(E-z_2)}} = {1\over{z_1 - z_2}} [ {1\over{E-z_1}}  
- {1\over{E-z_2}} ]
\end{equation}
\noindent
and we can write the resonant part (4.10) of the partial wave  
amplitude as the sum of two Breit-Wigner amplitudes

\begin{equation}
T_{res} = C^{(2)}_1 {{-{1\over 2} \Gamma_1}\over{E-E_1 + {1\over 2}  
i\Gamma_1}} + C^{(2)}_2 {{-{1\over 2} \Gamma_2}\over{E-E_2 +  
i\Gamma_2}}
\end{equation}
\noindent
where the complex coefficients
\begin{equation}
C^{(2)}_1 = 1 - 2i {{{1\over 2} \Gamma_2}\over{z_1 - z_2}}
\end{equation}

\[
C^{(2)}_2 = 1 + 2i {{{1\over 2} \Gamma_1}\over{z_1 - z_2}}
\]
\noindent
are exactly such that the unitarity condition
\begin{equation}
|e^{2i\delta^r} | = | 1 + 2i T_{res} | = 1
\end{equation}
\noindent
is satisfied for all $E$. The energy dependence of the widths  
$\Gamma_n (E)$ introduces energy dependence in $C_n(E), n=1,2$.

Consider now the case of three interfering resonances $N=3$. Then
\begin{equation}
T_{res} = a_1 + a_2 + a_3 + 2i (a_1 a_2 + a_1 a_3 + a_2 a_3 ) +  
(2i)^2 a_1 a_2 a_3
\end{equation}
\noindent
We can write the last term as a sum
\begin{equation}
{1\over{(E-z_1) (E-z_2)(E-z_3)}} = [{A\over{E-z_1}} +  
{B\over{E-z_2}} + {C\over{E-z_3}}] {1\over D}
\end{equation}
\noindent
Requiring that the terms proportional to $E^2$ and $E$ in the  
numerator on r.h.s. of (4.20) vanish, we obtain a sum
\begin{equation}
{K_1\over{E-z_1}} + {K_2\over{E-z_2}} + {K_3\over{E-z_3}}
\end{equation}
\noindent
where
\begin{equation}
K_1 = {A\over D} = {{z_3 - z_2}\over X}
\end{equation}
\[
K_2 = {B\over D} = {{z_3 - z_1}\over X}
\]
\[
K_3 = {C\over D} = {{z_1-z_2}\over X}
\]
\noindent
with
\begin{equation}
X = z_1 z_2 (z_1 - z_2) + z_3 z_1 (z_3 - z_1) + z_3 z_2 (z_3 - z_2)
\end{equation}
\noindent
The resonant part of the partial wave amplitude is again a coherent  
sum of the Breit-Wigner terms with complex coefficients
\begin{equation}
T_{res} = \sum\limits^3_{n=1} C^{(3)}_n {{-{1\over 2}  
\Gamma_n}\over{E - E_n + {1\over 2} i \Gamma_n}}
\end{equation}
\noindent
where
\begin{equation}
C^{(3)}_1 = 1 - 2i {{{1\over 2} \Gamma_2}\over{z_1 - z_2}} - 2i  
{{{1\over 2} \Gamma_3}\over{z_1 - z_3}} + (2i)^2 ({1\over  
2}\Gamma_2) ({1\over 2} \Gamma_3)K_1
\end{equation}
\[
C^{(3)}_2 = 1 + 2i {{{1\over 2} \Gamma_2}\over{z_1 - z_2}} - 2i  
{{{1\over 2} \Gamma_3}\over{z_2 - z_3}} + (2i)^2 ({1\over  
2}\Gamma_1) ({1\over 2} \Gamma_3)K_2
\]
\[
C^{(3)}_3 = 1 + 2i {{{1\over 2} \Gamma_2}\over{z_1 - z_3}} + 2i  
{{{1\over 2} \Gamma_2}\over{z_2 - z_3}} + (2i)^2 ({1\over  
2}\Gamma_1) ({1\over 2} \Gamma_2)K_3
\]
This procedure is general and valid for any finite $N$. Assuming  
that the resonant phase $\delta^r$ can be separated from the phase  
shift $\delta$ and is given by the sum of Breit-Wigner phases, we  
will always get the resonant part $T_{res}$ of the partial wave  
amplitudes $T^I_\ell$ in (4.9) as a sum of Breit-Wigner amplitudes
\begin{equation}
T_{res} (E) = \sum\limits^N_{n=1} C^{(N)}_n (E) {{{1\over 2}  
\Gamma_n (E)}\over{E- E_n + {1\over 2} i \Gamma_n (E)}}
\end{equation}
\noindent
In (4.26) the complex coefficients $C^{(N)}_n$ have an explicit  
form in terms of resonance parameters $E_n, \Gamma_n, n=1,\ldots ,  
N$ such that $T_{res}$ satisfies the unitary condition (4.18). The  
form of coefficients $C^{(N)}_n$ depends on the number of resonances  
contributing to the partial wave $T^I_\ell$.

As the result of (4.26) we can conclude that the multiresonance  
parametrization of partial wave amplitudes based on additivity of  
Breit-Wigner phases has a general form of a coherent sum of  
Breit-Wigner amplitudes $a_n$ with complex coefficients and a  
complex coherent background
\begin{equation}
T = {1\over q} [B(E) + \sum\limits^N_{n=1} R^{(N)}_n (E) {{-{1\over  
2} \Gamma_n}\over{E - E_n + i {1\over 2} \Gamma_n}} ]
\end{equation}
\noindent
where
\begin{equation}
B = {1\over{2i}} [ \eta e^{2i\xi} - 1]
\end{equation}
\[
R^{(N)}_n = e^{2i\xi} \eta C^{(N)}_n = (1+2iB) C^{(N)}_n
\]
\noindent
Comparing (4.27) with expression (2.19), we see that the  
description of multiresonance contributions using the addition of  
Breit-Wigner phases leads to the same form of partial wave  
amplitudes as the analyticity of the $S$-matrix. However, the  
complex background $B$ and the complex coefficients $R^{(N)}_n$ in  
(4.27) have the explicit form (4.28) imposed by the additivity of  
Breit-Wigner phases.

Note that in the derivation of (4.26) for $T_{res}$, and in the  
resultant form (4.27) with (4.28), we have not needed or used the  
assumption that inelasticity $\eta \le 1$. The Hu method is based on  
the unitarity of $1+2i T_{res}$ and is not related to the unitarity  
of the whole partial wave amplitude $T^I_\ell$.

Finally we give a relativistic form for the multiresonance  
description of partial wave amplitudes. The relativistic form of  
Breit-Wigner amplitudes (4.7) is given by
\begin{equation}
a_n = {{-m_n \Gamma_n (s)}\over{s^2 - m^2_n + i m_n \Gamma_n (s)}}
\end{equation}
\noindent
where we have used $m_n$ instead of $E_n$ to emphasize that $E_n$  
is the mass of the resonance. To obtain the corresponding  
coefficients $C^{(N)}_n$, we make replacements in (4.17) or (4.25)
\begin{equation}
{1\over 2} \Gamma_n \to m_n \Gamma_n
\end{equation}
\[
z_n = E_n - {1\over 2} i \Gamma_n \to z_n = m^2_n - i m_n \Gamma_n
\]
\noindent
The partial wave amplitudes then have the relativistic form
\begin{equation}
T(s) = {1\over q} [B(s) + \sum\limits^N_{n=1} R^{(N)}_n (s) a_n (s) ]
\end{equation}
\noindent
where $B$ and $R^{(N)}_n$ are still given by (4.28) with  
replacements (4.30) to satisfy the unitarity of $T_{res}$.

\section{Generalized dispersion relations for partial wave  
amplitudes and interfering resonances in $\pi\pi$ scattering.}

In this section we shall relate the multiresonance parametrization  
(4.31) of partial wave amplitudes $T^I_\ell$ with a multiresonance  
parametrization obtained from analyticity. To this end we shall use  
generalized dispersion relations for the amplitudes
\begin{equation}
t^I_\ell (s) = q T^I_\ell (s)
\end{equation}
\noindent
where $s$ is the Mandelstam energy variable.

Our starting point is the well-known\cite{cush75} dispersion  
respresentation of a complex function $f(z)$ with simple poles at  
$z_n, n=1,2,\ldots , N$ in the complex plane $z$, a branch cut along  
a positive real axis from $\alpha$ to $\infty$ and with asymptotic  
property $|z|f(z) \to 0$ as $|z|\to\infty$. We shall also assume  
that the function $f(z)$ is a real function $f(z^*) = f^*(z)$. Using  
Cauchy's integral theorem and the process of contour deformation,  
it can be shown\cite{cush75} that
\begin{equation}
f(z) = \sum\limits^N_{n=1} {R_n\over{z-z_n}} + {1\over\pi}  
\int\limits^\infty_\alpha {{Im f(x^\prime) dx^\prime}\over{x^\prime  
- z}}
\end{equation}
\noindent
A remarkable feature of the proof of (5.2) is that it takes place  
for a fixed value of $z$\cite{cush75}. As the result, the dispersion  
relation (5.2) is also valid for moving poles for which $z_n = z_n  
(z)$. In such a case the residues $R_n$ in (5.2) also depend on $z$,  
i.e. $R_n = R_n (z)$. Furthermore, the dispersion relation (5.2) is  
easily generalized to include a left-hand cut and for functions  
that are not real. In the latter case $Im f(x^\prime)$ in (5.2) is  
replaced by a discontinuity function along the cut(s).

In $\pi\pi$ scattering, the partial wave amplitudes $t^I_\ell(s)$  
have a right-hand cut for $s \ge 4\mu^2$ (where $\mu$ is the pion  
mass), and a left-hand cut for $s\le 0$ due to Mandelstam  
analyticity\cite{gasio66}. Let us assume that the amplitude  
$t^I_\ell$ has a finite number $N^I_\ell$ of complex poles
\begin{equation}
s_n = m^2_n - i m_n \Gamma (s),\ n=1, \ldots , N^I_\ell
\end{equation}
\noindent
corresponding to the resonances in $t^I_\ell$. Note that the  
imaginary part of the poles depends on the energy variable $s$. In  
principle, the mass $m_n$ could also depend on the energy $s$. This  
possibility has been recently considered by Pennington\cite{penn99}.  
Omitting the indices $I$ and $\ell$, the generalized dispersion  
relations for the partial wave amplitude $t^I_\ell$ read
\begin{equation}
t(s) = I(s) + \sum\limits^N_{n=1} {{R_n(s)}\over{s-s_n (s)}}
\end{equation}
\noindent
where $I(s)$ are the dispersion integrals over the left-hand and  
right-hand cuts\cite{gasio66} and $R_n(s)$ are the pole residues. It  
is convenient to rewrite (5.4) in a form using Breit-Wigner  
amplitudes $a_n(s)$
\begin{equation}
t(s) = I(s) + \sum\limits^N_{n=1} R_n(s) a_n (s)
\end{equation}
\noindent
where we have redefined the pole residues with
\begin{equation}
a_n(s) = {{-m_n \Gamma_n(s)}\over{s-m^2_n + im_n \Gamma_n(s)}}
\end{equation}
\noindent
The representation (5.5) is valid for all $s\ge 4\mu^2$. The  
representation (5.5) of partial waves $t^I_\ell$ coincides with the  
parametrization (4.31) provided that
\begin{equation}
I(s) \equiv B(s) = {1\over{2i}} [\eta (s) e^{2i\xi(s)} -1]
\end{equation}
\[
R_n (s) \equiv R^{(N)}_n (s) = \eta (s) e^{2i\xi(s)} C^{(N)}_n (s)
\]
\[
= (1+ 2i B) C^{(N)}_n
\]
\noindent
We see that the multiresonance parameterization based on additivity  
of Breit-Wigner phases (4.4) imposes a special form on the  
dispersion integrals and pole residues given by (5.7).

In general, a partial wave $t^I_\ell$ can be written in two forms
\begin{equation}
t = {1\over{2i}} [\eta e^{2i\delta} - 1] = I + \sum\limits^N_{n=1}  
R_n a_n
\end{equation}
\noindent
Apart from the partial wave unitarity equations (3.1) and (3.4) and  
the analyticity assumptions, there are no constraints on the  
partial waves. The assumption of additivity of Breit-Wigner phases  
(4.4) is a new constraint that restricts the partial waves to  
analytical functions that satisfy the conditions (5.7). We find no  
physical justification for such a restriction and no advantage in  
using it in phenomenological fits to data to determine resonance  
parameters.

\section{Interfering resonances in a finite energy interval.}

In the previous two sections we have assumed that $N$ is the total  
number of resonances contributing to a partial wave. The  
parametrizations (4.31) and (5.5) were valid for all energies  $s  
\ge 4\mu^2$. In practice $N$ is not known and fits to data are done  
in a finite energy interval. Such is the case e.g. of analyses  
\cite{bugg96,ishida96}. In this Section we develop parametrizations  
of partial waves appropriate for analyses in a finite energy  
interval where only a few resonances contribute. The  
parametrizations will be based both on additivity of Breit-Wigner  
phases and analyticity and we shall compare their use in practical  
fits to data.  The results will be used in the Section VIII.

Let us consider energy interval $4\mu^2 \le s \le s_M$ where $M$  
resonances contribute. In the framework of the assumption of  
additivity of Breit-Wigner phases we will assume that the resonant  
phases of resonances outside of this energy interval are absorbed in  
the background phase. The total phase shift then is
\begin{equation}
\delta = \xi^{(M)} + \delta^r_M
\end{equation}
\noindent
where
\begin{equation}
\delta^r_M = \sum\limits^M_{n=1} \delta^r_n
\end{equation}
\begin{equation}
\xi^{(M)} = \xi + \sum\limits^N_{n=M+1} \delta^r_n
\end{equation}
\noindent
The partial wave then takes the form
\begin{equation}
t = {1\over{2i}} (\eta e^{2i\xi^{(M)}} - 1) + \eta e^{2i\xi^{(M)}}  
T^{(M)}_{res}
\end{equation}
\noindent
where the resonant part
\begin{equation}
T^{(M)}_{res} = \sum\limits^M_{n=1} C^{(M)}_n a_n
\end{equation}
\noindent
is unitary
\begin{equation}
e^{2i\delta^r_M} = 1 + 2i T^{(M)}_{res}
\end{equation}
\noindent
Alternatively we can rewrite (4.31) in the form
\begin{equation}
t = B^{(MN)} + \sum\limits^M_{n=1} R^{(N)}_n a_n
\end{equation}
\noindent
where
\begin{equation}
B^{(MN)} = {1\over{2i}} (\eta e^{2i\xi} - 1) + \sum\limits^N_{m =  
M+1} R^{(N)}_m a_m
\end{equation}
\noindent
is the background term. Note that the sum in (6.7) is not unitary.  
We cannot compare the coefficients of Breit-Wigner amplitudes $a_n,\  
n=1,\ldots , M$ in (6.4) with those in (6.7) since $\xi^{(M)}$ in  
(6.4) contains the terms $a_m,\ m=M+1, \ldots , N$ but the sum in  
(6.7) does not.

If we look at the general form (5.5) from analyticity, then for $s  
< s_M$ we can write
\begin{equation}
t=B^{(M)} + \sum\limits^M_{n=1} R_n a_n
\end{equation}
\noindent
where the background term
\begin{equation}
B^{(M)} = I + \sum\limits^N_{m = M+1} R_m a_m
\end{equation}
\noindent
In (6.9) the residues $R_n$ are not constrained by the conditions (5.7).

In fitting data using the parametrization (6.4) we explicitly make  
use of the assumption of additivity of Breit-Wigner phases. This is  
also the case when we use (6.7) if $N$ is known and the coefficients  
$C^{(N)}_n$ in $R^{(N)}_n$ can be calculated. In general $N$ is not  
known, and the background $B^{(MN)}$ and residues $R^{(N)}_n$ are  
free parameters. Then there is no difference in using (6.7) or the  
general form (6.9) from analyticity alone, since in (6.9) the  
background $B^{(M)}$ and residues $R_n$ are not constrained except  
for unitarity. In all cases we use constrained optimization of the  
$\chi^2$ function. In the case of (6.4) we require that inelasticity  
function $\eta \le 1$. In the case of (6.7) or (6.9) we require  
that $Im t \ge |t|^2$ and use programs such as MINOS\cite{murtagh83}  
for constrained optimization.

It is not obvious that the use of parametrization (6.4) from  
additivity of Breit-Wigner phases and the parametrization (6.9) from  
analyticity alone, will lead to the same resonance parameters in  
both cases. The use of parametrization (6.4) confers no  
phenomenological or computational advantage over the parametrization  
(6.9). The assumption of additivity of Breit-Wigner phases  
restricts the background and the complex coefficients multiplying  
the Breit-Wigner amplitudes $a_n,\ n=1, \ldots , M$ in the  
parametrization (6.4) to specific forms. Since there is no physical  
justification for such a restriction and the parametrization (6.9)  
is free from such constraints, we suggest that the use of  
parametrization (6.9) is more appropriate in determining resonance  
parameters in $\pi\pi$ scattering.

\section{Unitarity in pion production $\pi^-  \lowercase{p}  \to  
\pi^- \pi^+ \lowercase{n}$}

It is a common misconception to identify the partial wave  
production amplitudes in reaction $\pi^- p \to \pi^- \pi^+ n$ with  
partial waves $T^I_\ell$ in $\pi\pi$ scattering and demand that the  
partial wave production amplitudes also satisfy the partial wave  
unitarity (3.1). In this Section we clarify the distinction between  
the two kinds of amplitudes and the associated unitarity relations.

The production process $\pi^- p \to \pi^- \pi^+ n$ is described by  
production
 amplitudes\cite{martin76,lutz78,svec92}
\begin{equation}
T_{\lambda_n, 0\lambda_p}(s,t,m^2,\theta,\phi)
\end{equation}
 where $\lambda_p$ and $\lambda_n$ are proton and neutron  
helicities, s is the c.m.s. energy squared, t is the momentum  
transfer between the incident pion and the dipion system $(\pi^-  
\pi^+)$, $m^2$ is the dipion mass squared, and  
$\Omega=(\theta,\phi)$ is the solid angle of the final $\pi^-$ pion  
in the dipion rest frame. The dipion state does not have a definite  
spin. The production amplitudes (7.1) can be expressed in terms of  
partial wave production amplitudes $M^J_{\lambda,\lambda_n,  
0\lambda_p}(s,t,m^2)$ corresponding to definite dimeson spin $J$  
using the angular expansion\cite{martin76,lutz78,svec92}

\begin{equation}
T_{\lambda_n, 0\lambda_p} = \sum\limits^\infty_{j=0}  
\sum\limits^{+J}_{\lambda=-J} (2J+1)^{1/2} M^J_{\lambda \lambda_n,  
0\lambda_p}(s,t,m^2) d^J_{\lambda 0}(\theta) e^{i\lambda\phi}
\end{equation}

\noindent
where $J$ is the spin and $\lambda$ the helicity of the  
$(\pi^-\pi^+)$ dimeson system.

It is evident from (7.2) that the partial wave production  
amplitudes $M^J_{\lambda,\lambda_n, 0\lambda_p}(s,t,m^2)$ cannot be  
identified with the $\pi\pi$ partial wave amplitudes $T^I_J(m^2)$.  
The amplitudes $M^J_{\lambda,\lambda_n, 0\lambda_p}(s,t,m^2)$ can be  
thought of as two-body helicity amplitudes for a process $\pi^- + p  
\to M(J,m) + n$ where the "particle" $M(J,m)$ has spin J and mass  
$m$.

The production amplitudes $T_{\lambda_n, 0\lambda_p}$ satisfy the  
unitarity condition
\cite{pilk67}

\begin{equation}
-i(T_{\lambda_n, 0\lambda_p} - T^*_{0\lambda_p,\lambda_n}) =  
\sum\limits_n \int{T_{0\lambda_p,n}T^*_{\lambda_n,n}d\Phi_n}
\end{equation}

\noindent
where $d\Phi_n$ is the $n$-body Lorentz invariant phase space of  
the intermediate state $n$.
Since the initial state in $\pi^- p \to \pi^- \pi^+ n$ is a  
two-body state and the final state is a three-body state, the  
amplitude   $T_{\lambda_n, 0\lambda_p}$ enters the unitarity  
integral only linearly. This occurs only when the intermediate state  
is $\pi^- p$ or $\pi^-\pi^+n$. However the three-body intermediate  
state involves a $3 \to 3$ amplitude and a three-body phase space  
integral. Separating the two-body intermediate states $\pi^-p$ and  
$\pi^0n$, we can write (7.3) in the form

\begin{equation}
-i(T_{\lambda_n, 0\lambda_p} - T^*_{0\lambda_p,\lambda_n}) =  
\sum\limits_{\lambda'_p}\int{T_{0\lambda_p,  
0\lambda'_p}T^*_{\lambda_n, 0\lambda'_p}d\Phi_2} +
\sum\limits_{\lambda'_n}\int{T_{0\lambda_p,  
0\lambda'_n}T^*_{\lambda_n, 0\lambda'_n}d\Phi_2} +
\Delta_{\lambda_n,0\lambda_p}
\end{equation}

\noindent
where $T_{0\lambda_p,0\lambda'_p}$ and   
$T_{0\lambda_p,0\lambda'_n}$ are helicity amplitudes of reactions  
$\pi^- p \to \pi^- p$ and $\pi^- p \to \pi^0 n$, respectively. The  
amplitude $T^*_{\lambda_n, 0\lambda'_n}$ corresponds to process  
$\pi^0 n \to \pi^-\pi^+ n$. The inelastic unitarity contribution  
$\Delta_{\lambda_n,0\lambda_p}(s,t,m^2,\theta,\phi)$ can be expanded  
in the form analogous to (7.2)

\begin{equation}
\Delta_{\lambda_n, 0\lambda_p} = \sum\limits^\infty_{j=0}  
\sum\limits^{+J}_{\lambda=-J} (2J+1)^{1/2} \Delta^J_{\lambda  
\lambda_n, 0\lambda_p}(s,t,m^2) d^J_{\lambda 0}(\theta)  
e^{i\lambda\phi}
\end{equation}

\noindent
Using expansions (7.2) and (7.5) in (7.4) we get unitarity  
relations for partial wave production amplitudes

\begin{equation}
-i(M^J_{\lambda \lambda_n, 0\lambda_p} - M^{J*}_{0\lambda_p,\lambda  
\lambda_n}) = \sum\limits_{\lambda'_p}\int{T_{0\lambda_p,  
0\lambda'_p}M^{J*}_{\lambda_n, 0\lambda'_p}d\Phi_2} +
\sum\limits_{\lambda'_n}\int{T_{0\lambda_p,  
0\lambda'_n}M^{J*}_{\lambda_n, 0\lambda'_n}d\Phi_2} +
\Delta^J_{\lambda_n,0\lambda_p}
\end{equation}

\noindent
Using time-reversal relations for two-body  helicity  
amplitudes\cite{bourrely80}

\begin{equation}
M^J_{0\lambda_p,\lambda \lambda_n} = (-1)^{\lambda_n - \lambda_p -  
\lambda}M^J_{\lambda\lambda_n, 0\lambda_p}
\end{equation}

\noindent
 we see that the left hand side of the partial wave unitarity  
relation (7.6) does not simplify to $2ImM^J_{\lambda\lambda_n,  
0\lambda_p}$ as is the case for the partial waves $T^I_\ell$ in  
$\pi\pi$ scattering. The right hand side of (7.6) involves  
$M^J_{\lambda\lambda_n, 0\lambda_p}$ only linearly and not  
quadratically as is the case in $\pi\pi$ scattering. Futhermore, the  
right hand side of unitarity relation (7.6) includes (linearly)  
partial wave production amplitudes for the process $\pi^0 n \to  
\pi^- \pi^+ n$. We conclude that the unitarity relations (7.6) for  
partial wave production amplitudes $M^J_{\lambda\lambda_n,  
0\lambda_p}$
are complex relations that do not have the simple form

\begin{equation}
Im T^I_\ell = q |T^I_\ell|^2 + \Delta^I_\ell
\end{equation}

\noindent
of the partial wave unitarity relations in $\pi\pi$ scattering.

For brevity let us define $M^J_{\Lambda} \equiv  
M^J_{\lambda\lambda_n, 0\lambda_p}$ where $\Lambda$ stands for the  
helicities. The amplitude $M^J_{\Lambda}$ is a complex function and  
so is the function $1 + 2iqM^J_{\Lambda}$. In analogy with (3.12)  
and (3.3) we can write

\begin{equation}
M^J_{\Lambda} = {1\over{2iq}}(\eta^J_{\Lambda}  
e^{2i\delta^J_{\Lambda}} -1)
\end{equation}

\noindent
where $\eta^J_{\Lambda}(s,t,m^2)$ is the "inelasticity" and  
$\delta^J_{\Lambda}(s,t,m^2)$ is the "phase shift". The amplitude  
$M^J_{\Lambda}$ satisfies relation similar to  (3.13)

\begin{equation}
ImM^J_{\Lambda} = q|M^J_{\Lambda}|^2 + {1\over{4q}}(1 -  
(\eta^J_{\Lambda})^2)
\end{equation}

\noindent
Unlike in $\pi\pi$ scattering, the form of the equation (7.10) does  
not coincide with the form of partial wave unitarity (7.6) and the  
"inelasticity" $\eta^J_{\Lambda}$ cannot be related to the inelastic  
unitarity contributions $\Delta^J_{\Lambda}$, in contrast to (3.4).

\section{Interfering resonances in production processes.}

The amplitudes describing the production processes such as $\pi N  
\to \pi\pi N$, $pp\to \pi\pi pp$ or $p\overline p \to 3\pi$ are far  
more complex than the isospin amplitudes in the $\pi\pi$ scattering.  
As an example, consider pion production in $\pi^- p \to \pi^- \pi^+  
n$. The angular distribution of the dipion $\pi^- \pi^+$ state is  
described by partial wave production amplitudes  
$M^J_{\lambda\lambda_n, 0\lambda_p} (s,t,m^2)$ defined in the  
previous Section with eq. (7.2). The measurements of $\pi^- p \to  
\pi^- \pi^+ n$ on polarized target actually determine the moduli of  
nucleon transversity amplitudes \cite{lutz78,svec92} which are  
linear combinations of helicity amplitudes $M^J_{\lambda\lambda_n,  
0\lambda_p} $. For masses $m$ \appls 1000 MeV, the pion production  
is described by two $S$-wave $(J=0)$ nucleon transversity amplitudes  
$S$ and $\overline S$, and by six $P$-wave $(J=1)$ nucleon  
transversity amplitudes $L, \overline L, N, \overline N, U,  
\overline U$\cite{lutz78,svec92}. The amplitudes $\overline A =  
\overline S, \overline L, \overline N, \overline U$ correspond to  
nucleon transversity ``up'' while the amplitudes $A = S, L, N, U$  
correspond to nucleon transversity ``down''. The amplitudes  
$L,\overline L$ correspond to dimension helicity $\lambda=0$ while  
$N, \overline N$ and $U, \overline U$ are natural and unnatural  
parity amplitudes corresponding to combinations of $\lambda = \pm  
1$.

The measurements of pion production on polarized targets enable to  
advance hadron spectroscopy from the level of spin-averaged  
cross-sections to the level of spin-dependent production amplitudes.  
These measurements determine the mass spectra  $|A|^2$ and  
$|\overline A|^2$ of spin-dependent  production amplitudes.  
Measurements of $\pi^- p \to \pi^- \pi^+ n$ on transversely  
polarized targets were done at CERN at
17.2 GeV/c\cite{becker79,becker79b,chabaud83,rybicki85} and at ITEP at 
1.78 GeV/c\cite{alek99}. Measurements of
$\pi^+ n \to \pi^+\pi^- p$\cite{svec92,lesquen85,svec96,svec97a} and
$K^+ n \to K^+\pi^- p$\cite{lesquen89,svec92b} at 5.98 and 11.85  
GeV/c on transversely polarized deuteron target were also done at  
CERN. More recently it has been shown that mass spectra of  
production amplitudes can be obtained in measurements of
$\pi^- p \to \pi^0\pi^0 n$, $\pi^- p \to \eta\eta n$\cite{svec97b}  
and $\pi^- p \to \eta\pi^- n$,
$\pi^- p \to \eta\pi^0 n$\cite{svec97c} on transversely polarized  
targets allowing for amplitude spectroscopy of these interesting  
processes.

The analysis of mass spectra measured in production processes  
requires a parametrization of the production amplitudes in terms of  
the Breit-Wigner amplitudes to identify contributing resonances and  
to determine their parameters.  Here we discuss two approaches, one  
based on the additivity of Breit-Wigner phases and the other on  
analyticity of production amplitudes $A(s,t,m^2)$ in the mass  
variable at fixed $s$ and $t$.

First we note that the unitarity equation (7.6) for the partial  
wave production amplitudes in $\pi N \to \pi\pi N$ is a complex  
relation and that the helicity amplitudes $M^J_{\Lambda}$ or the   
transversity amplitudes $A$ and $\overline A$ do not satisfy the  
two-body partial wave unitarity equation (3.1) with (3.4).  
Nevertheless, the experimentally measured production amplitudes  
$A(s,t,m^2)$ are complex functions and as such can be written in the  
form
\begin{equation}
A (s,t,m^2) = {1\over{2i}} [\eta_A e^{2i\delta_A} - 1]
\end{equation}
\noindent
where the ``inelasticity'' $\eta_A = \eta_A (s,t,m^2)$ and ``phase  
shift'' $\delta_A = \delta_A (s,t,m^2)$ depend also on the  
helicities or transversities of the amplitude $A$. Obviously,
\begin{equation}
Im A = |A|^2 + {1\over 4} (1-\eta^2_A)
\end{equation}
\noindent
However, there is no requirement now that $\eta_A \le 1$ since  
$\eta_A$ has no relation to unitarity as in (3.4).

We can pursue the analogy with the $\pi\pi$ scattering, and impose  
an assumption that the ``phase shift''
\begin{equation}
\delta_A (s,t,m^2) = \xi_A (s,t,m^2) + \delta^r (m^2)
\end{equation}
\noindent
where $\delta^r$ is the sum of Breit-Wigner phases of the $N$  
resonances contributing to the amplitude $A$ and $\xi_A$ is the  
``background'' phase. If we restrict ourselves to a finite mass  
interval $4\mu^2 \le m^2 \le m^2_M$ with $M$ resonances, we can  
write
\begin{equation}
\delta_A (s,t,m^2) = \xi^{(M)}_A (s,t,m^2) + \delta^r_M (m^2)
\end{equation}
\begin{equation}
A(s,t,m^2) = {1\over{2i}} (\eta_A e^{2i\xi_A^{(M)}} - 1) + \eta_A  
e^{2i\xi^{(M)}_A} T^{(M)}_{res}(m^2)
\end{equation}
\noindent
in analogy with (6.4) for $\pi\pi$ scattering.

A more general approach is to use analyticity of $A(s,t,m^2)$ in  
$m^2$ with $s$ and $t$ fixed. We can assume that kinematical  
singularities have been removed from the production amplitudes  
$A(s,t,m^2)$\cite{martin70}. Assuming that there are $N$  
Breit-Wigner poles in the amplitude $A(s,t,m^2)$ in the mass  
variable $m^2$, we can use the generalized dispersion relations for  
the variable $m^2$ with $s$ and $t$ fixed to get
\begin{equation}
A(s,t,m^2) = I (s,t,m^2) + \sum\limits^N_{n=1} R_n(s,t,m^2) a_n (m^2)
\end{equation}
\noindent
where $I$ is the contribution of dispersion integrals, $R_n$ are  
complex pole residues, and $a_n$ are the Breit-Wigner amplitudes  
(5.6). In a finite mass interval $4\mu^2 \le m^2 \le m^2_M$ with $M$  
resonances we can write
\begin{equation}
A(s,t,m^2) = B^{(M)} (s,t,m^2) + \sum\limits^M_{n=1} R_n (s,t,m^2)  
a_n (m^2)
\end{equation}
\noindent
where the background
\begin{equation}
B^{(M)}(s,t,m^2) = I + \sum\limits^N_{m=M+1} R_m a_m
\end{equation}
We note that for $M=N$ we get back the constraints (5.7) with  
replacements $\eta \to \eta_A$ and $\xi \to \xi_A$. Again, the  
assumption of additivity of Breit-Wigner phases restricts the  
production amplitudes to analytical functions that satisfy the  
constraints (5.7).

The measured mass spectra $|A|^2$ can now be fitted either with the  
parametrization (8.5) or with the more general parametrization  
(8.7). There are no unitarity constraints to be imposed on the  
production amplitudes $|A|^2$ during the fits since the right hand  
side of the unitarity relation (7.6) is not known and the partial  
wave unitarity (3.1) or (7.8) for $\pi\pi$ amplitudes  
$T^I_{\ell}(s)$ does not apply to the production amplitudes  
$A(s,t,m^2)$. The unitarity constraint (3.1) or (7.8) can be imposed  
only in the analysis of data on the $\pi\pi \to \pi\pi$ reaction  
and below we discuss its effect on $\pi\pi$ amplitudes.

 Since there is no physical justification for the assumption of  
additivity of Breit-Wigner phases in (8.3) and since the form (8.5)  
confers no phenomenological or computational advantage over the more  
general analytical form (8.7) , we conclude that the use of the  
form (8.7) is more appropriate in fits to mass spectra in production  
processes to determine the resonance parameters of interfering  
resonances.

The parametrization of production amplitudes in terms of a coherent  
sum with complex coefficients and a complex coherent background as  
in (8.7) has been an accepted practice for a long time. Such  
parametrizations first appeared in connection with the possible  
double-pole character of the $A_2$ meson\cite{lassila67} and the  
splitting of the $Q$ resonance in $K^+ \pi^-\pi^+$ mass  
spectrum\cite{barnham70}. Recently such parametrization has been  
used in the study of $\sigma(750) - f_0(980)$ interference in  
$S$-wave production amplitudes in $\pi^- p \to \pi^-\pi^+n$ measured  
on polarized target at CERN\cite{svec97,svec98} and in the study of  
$\sigma -f_0(980)$ interference in the central collision $pp\to  
\pi^0 \pi^0 pp$\cite{alde97}. More recently, an analysis of $S$-wave  
production amplitudes from threshold to 2 GeV in $pp\to \pi^0\pi^0  
pp$ was made using three\cite{barberis99} and four\cite{bella99}  
interfering Breit-Wigner amplitudes and a coherent background. The  
GAMS collaboration used four interfering Breit-Wigner amplitudes and  
a coherent background in their fit of $S$-wave mass spectrum from  
threshold to 3 GeV in $\pi^- p\to \pi^0\pi^0 n$ measured at 100  
GeV\cite{alde98}. Also recently, the Fermilab E791 Collaboration  
used the form (8.7) to fit Dalitz plot of $D^+ \to \pi^- \pi^+  
\pi^+$ decays in their search for a scalar resonance  
$\sigma$\cite{aitala01}.

Finally we comment on the determination of $\pi\pi$ partial wave  
amplitudes from measurements of $\pi N \to \pi\pi N$. The resonance  
parameters from the fits to mass spectra such as those measured in  
$\pi N \to \pi^+ \pi^- N$ on polarized  
targets\cite{svec97,svec98,svec97a} or in $\pi^- p \to \pi^0\pi^0  
n$\cite{barberis99,bella99,alde98} must be the same in $\pi\pi$  
partial waves. However, the $\pi\pi$ partial wave amplitudes are  
expected to satisfy the partial wave unitarity constraints (3.1) and  
(3.4), or rather the inequalities (3.8) and (3.9) which for the  
amplitudes $t^I_\ell$ defined in (5.1) read
\begin{equation}
|t^I_\ell|^2 \le Im t^I_\ell \le |t^I_\ell|^2 +{1\over{4}}
\end{equation}
\noindent
The unitarity conditions (8.9) can always be satisfied by an  
appropriate choice of background and complex residues $R_n(s)$ in  
the general parametrization (6.9) based on analyticity. Although the  
$\pi\pi$ partial waves and production amplitudes in $\pi N\to  
\pi\pi N$ with the same spin and isospin share the same Breit-Wigner  
poles, they are different analytical functions and thus the  
residues of the poles and the backgrounds are different. In  
particular, the residues in production amplitudes A depend on  
particle helicities and kinematic variables $s$ and $t$. Accepting  
the resonance parameters obtained from the fits to the mass spectra  
$|A|^2$ measured in $\pi N \to \pi\pi N$ to describe the resonances  
in $\pi\pi \to \pi\pi$ scattering, the effect of the unitarity  
conditions (8.9) is to constrain the residues $R_n(s)$ and the  
background term in the general parametrization (6.9) of the $\pi\pi$  
amplitudes.

It is also possible to use the resonance parameters determined from  
measurements of $\pi N\to \pi\pi N$ to calculate the resonant part  
$T^{(M)}_{res}$ and to define the $\pi\pi$ partial waves using the  
parametrization (6.4) with free background and inelasticity  
functions $\xi^{(M)} (s)$ and $\eta (s)$. The unitarity can be  
satisfied by imposing the condition $\eta \le 1$.

The unitarity constraints (8.9) may not uniquely determine the  
background and pole residues in the parametrization (6.9) from  
analyticity and the use of parametrization (6.4) from additivity of  
Breit-Wigner phases is questionable. We conclude that the resonance  
parameters determined from mesurements of $\pi N\to \pi\pi N$ alone  
may not determine the $\pi\pi$ partial wave amplitudes without  
additional assumptions or direct measurements of $\pi\pi \to \pi\pi$  
reactions.

\section{Summary.}

We have shown in the case of $\pi\pi$ scattering that the  
assumption of additivity of Breit-Wigner phases in a partial wave  
amplitude leads to a sum of Breit-Wigner amplitudes with complex  
coefficients and a coherent background (4.31). The coefficients have  
a specific form (4.28) in terms of resonance parameters of all  
contributing resonances. The form (4.31) is a special case of the  
general form (5.5) based on analyticity and it is not related to the  
unitarity property of partial waves ((3.1) and (3.4)). The  
claims\cite{bugg96,ishida96} that the additivity of Breit-Wigner  
phases provides the correct description of interfering resonances in  
$\pi\pi$ scattering are not justified since there is no physical  
reason why the Breit-Wigner poles must have the specific residues  
imposed by this assumption.  We found that the Breit-Wigner phases  
of interfering resonances are not necessarily additive. We suggest  
that the general form (5.5) from analyticity is more appropriate in  
fits to data. The unitarity conditions (8.9)  $|t^I_\ell|^2 \le Im  
t^I_\ell \le |t^I_\ell|^2 + {1\over{4}}$ can be effectively imposed  
using the modern methods of constrained  
optimization\cite{luenberger73,gill81,murtagh83}.

Mass spectra in production processes are described by production  
amplitudes. We used the case of $\pi N \to \pi^+ \pi^- N$ reaction  
to illustrate the complexity of production amplitudes. Specifically,  
the production amplitudes do not satisfy the two-body partial wave  
unitarity equation (3.1), depend of particle helicities and on  
several kinematic variables in addition to the invariant mass. We  
have used analyticity of production amplitudes in the invariant mass  
variable to justify the common  
practice\cite{lassila67,barnham70,svec97,svec98,alde97,barberis99,bella99,alde98,aitala01}  
of writing the production amplitudes as a coherent sum of  
Breit-Wigner amplitudes with free complex coefficients and a complex  
coherent background in fits to measured mass spectra to determine  
the resonance parameters of interfering resonances. The two-body  
unitarity constraints on the $\pi\pi$ partial wave amplitudes with  
the same resonances can be satisfied by an appropriate choice of  
complex residues of the contributing Breit-Wigner poles. This  
reflects the fact that the $\pi\pi$ partial wave amplitudes and  
production amplitudes while sharing the same resonances are  
different analytical functions.

\acknowledgements
I wish to thank Taku Ishida for helpful discussions and e-mail  
correspondence.

\end{document}